# Improved NSGA-II Based on a Novel Ranking Scheme

Rio G. L. D'Souza, K. Chandra Sekaran, and A. Kandasamy

**Abstract**— Non-dominated Sorting Genetic Algorithm (NSGA) has established itself as a benchmark algorithm for Multiobjective Optimization. The determination of pareto-optimal solutions is the key to its success. However the basic algorithm suffers from a high order of complexity, which renders it less useful for practical applications. Among the variants of NSGA, several attempts have been made to reduce the complexity. Though successful in reducing the runtime complexity, there is scope for further improvements, especially considering that the populations involved are frequently of large size. We propose a variant which reduces the run-time complexity using the simple principle of space-time trade-off. The improved algorithm is applied to the problem of classifying types of leukemia based on microarray data. Results of comparative tests are presented showing that the improved algorithm performs well on large populations.

**Index Terms**— NSGA, Multiobjective optimization, MOEA, Evolutionary algorithms, Pareto-optimal solutions, Non-domination.

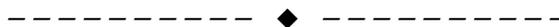

## 1 INTRODUCTION

MULTI-OBJECTIVE optimization has become a useful strategy in the solution of many modern engineering problems. Evolutionary algorithms have been found to be very successful in carrying out the optimization of multiple objectives. Most evolutionary algorithms are robust and multi-modal which proves to be a distinct advantage in the solution of such problems. Among the existing evolutionary algorithms for solving a Multiobjective optimization problem (MOOP) [1] the most prominent ones are the Pareto archived evolutionary strategy (PAES) [2], Strength pareto evolutionary algorithm (SPEA-2) [3] and the Non-dominated sorting genetic algorithm (NSGA-II) [4]. All these algorithms are based on the concept of Pareto-dominance.

NSGA-II has evolved over the last few years with many new variants which have attempted to reduce its time-complexity or improve its convergence to the true pareto front [5][6][7]. In this paper, we introduce a variant of NSGA-II which improves on the time-complexity by managing the book-keeping in a better way than the basic algorithm. For want of a better name, we call this variant the NSGA-IIa. We apply NSGA-IIa to the problem of classifying types of leukemia based on the gene expression dataset by Golub et al [8].

We also compare the performance of NSGA-IIa with that of the basic algorithm, NSGA-II, and also some of the other algorithms based on pareto-dominance that have been introduced in recent years. Results show that the new variant is on par with the other variants and hence can be considered in such situations which warrant an alternative approach for validation of results obtained by any of the other method.

## 2 RELATED WORK

Evolutionary algorithms attempt to mimic nature in the search for solutions to problems. In recent times they have been used to solve many real-world optimization problems. Early evolutionary algorithms were focused on optimizing single objectives. However, most optimization problems have multiple objectives. Optimization of multiple objectives requires that the relative importance of each objective be specified in advance which requires a prior knowledge of the possible solutions. But by using the concept of Pareto-dominance it is possible to avoid the need to know the possible solutions in advance [9]. This is one of the reasons for the popularity of such pareto-to-based approaches.

Multiobjective Evolutionary Algorithms (MOEAs) [10] are based for the following concepts:

1) Pareto-Front: The locus that is formed by a set of solutions that are equally good when compared to other solutions of that set is called as a Pareto-front.

2) Non-Domination: Non-dominated or pareto-optimal solutions are those solutions in the set which do not dominate each other, i.e., neither of them is better than the other in all the objective function evaluations. The solutions on each pareto-front are pareto-optimal with respect to each other.

Fonseca and Fleming [11] proposed one of the first multiobjective evolutionary algorithms called as Multiobjective genetic algorithm (MOGA). Knowles and Corne [2] introduced the PAES wherein the parent and offspring as well as the archived best solutions thus far are compared using pareto-dominance. Zitzler and Thiele [3] developed the SPEA-2, wherein the best solutions thus far are stored and compared using pareto-dominance with the current population.

―――――――――――

- *Rio G. L. D'Souza is with the Department of Computer Science and Engineering, St. Joseph Engineering College, Mangalore, India.*
- *K. Chandra Sekaran is with the Department of Computer Engineering, National Institute of Technology Karnataka, Surathkal, Mangalore, India.*
- *A. Kandasamy is with the Department of Mathematical and Computational Sciences, National Institute of Technology Karnataka, Surathkal, Mangalore, India.*

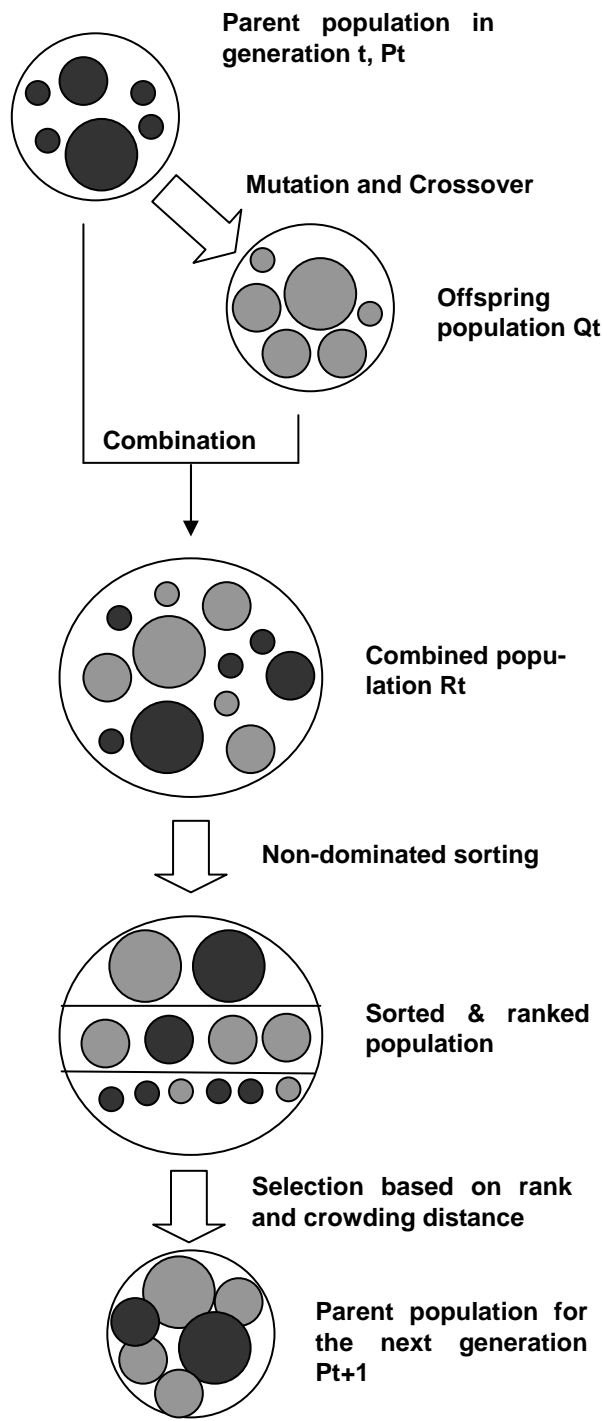

Fig. 1. Processing of populations in NSGA-II over one generation.

But the NSGA which was first proposed by Srinivas and Deb [12] in 1995 proved to be a landmark in the history of MOEAs. The time-complexity of the simple NSGA is $O(MN^3)$ where M is the number of objectives and N is the size of the dataset. Soon after, Deb and his students [4] developed a faster variant of NSGA, called NSGA-II, whose time-complexity is $O(MN^2)$.

During the last few years several researchers have come up with variants of NSGA-II which have a time-complexity of $O(MN \log_{M-1} N)$. The works by Fang [5] and Tran [6] are worth mentioning. Kumar et al. [13] have developed a memetic version of the NSGA-II. Jensen [7] has analyzed the techniques that could be used to improve the runtime of MOEAs and has also proposed an improved variant of NSGA-II. We compare our variant with this variant by Jensen (which we call as NSGA-IIb) in the Results section of this paper.

## 3 NON-DOMINATED SORTING GENETIC ALGORITHM

First we describe the working of the NSGA-II algorithm as given in [4]. Our variant is different only in the way the non-dominated sorting is performed. This variation we discuss in the next section. In the following discussion, we use the terms solution and individual to mean the same thing, since individuals in the population represent solutions to the problem that is being optimized.

In NSGA-II, we first create the offspring population Qt (of size N) using the parent population Pt (of size N), as shown in Fig. 1. The usual genetic operators such as single-point crossover and bit-wise mutation operators are used in this process. Next, we combine the two populations to form an intermediate population Rt of size 2N. Thereafter, we evaluate the fitness of each offspring in the 2N population using the multiple objective functions.

At this stage, we carry out non-dominated sorting procedure over the 2N population to rank and divide the individuals into different non-dominated fronts. The details of the non-dominated sorting and rank assignment are given in Fig. 2. Thereafter, we create the new parent population Pt+1 by choosing individuals of the non-dominated fronts, one at a time. We choose the individuals of best ranked fronts first followed by the next-best and so on, till we obtain N individuals.

Since the intermediate population Rt has a size of 2N, we discard those fronts which could not be accommodated. In case there is space only for a part of a front in the new population, we use a crowded-distance operator to determine the individuals among those in the front that are from the least crowded regions. We choose such individuals so as to fill up the required number in the new population Pt+1. Details of crowed-distance operator can be obtained from [7].

The complete NSGA-II procedure is given below:

**BEGIN**
While generation count is not reached
**Begin Loop**
• Combine parent Pt and offspring population Qt to obtain population Rt of size 2N.
• Perform Non-dominated Sort on Rt and assign ranks to each pareto front with fitness Fi.
• Starting from Pareto front with fitness F1, add each pareto-front Fi to the new parent population Pt+1 until a complete front Fi cannot be included.
• From the current pareto-front Fi, add individual members to new parent population Pt+1 until it reaches the size N.
• Apply selection, crossover and mutation to new par-

ent population Pt+1 and obtain the new offspring population Qt+1.
• Increment generation count.
**End Loop**
**END.**

As stated in [4], this algorithm has a runtime complexity of $O(MN^2)$ which is obvious from Fig. 2. In this figure, $S_p$ represents a set of solutions that the solution $p$ dominates and $n_p$ represents the domination count (the number of solutions which dominate the solution $p$). The symbol $\prec$ indicates domination ($p \prec q$ indicates that $p$ dominates $q$) and $F$ represents the pareto-front of solutions.

In Jensen [7], an improved version of this algorithm is presented. In this version, NSGA-IIb, all the pareto-fronts are constructed simultaneously by sweeping the solutions one-by-one in a way that guarantees that if sj solution is swept after solution si, then sj cannot dominate si. This is achieved by presorting all of the solutions on the objective values in such a way that the above condition will always hold if i < j.

Due to the presorting, which can be performed in $O(N \log N)$ time, this algorithm has a runtime complexity of $O(MN \log_{M-1} N)$. For the rest of the details of this method, please refer to [7].

## 4 THE IMPROVED NSGA-II

In our work, we have used a variation of the non-dominated sorting procedure shown in Fig. 3. As seen in the figure, we perform sorting of individuals based on each of the objectives, one after the other, till all objectives are considered.

During this sort, we keep track of the index of each individual, so that we know the position value of any given individual i in each sorted array. This information is critical since it helps us to rank the fronts in the next step.

We assign the rank of each individual by summing up the position value of that individual in all the objectives. Since similar position values where assigned to individuals having similar objective values, the sum of the position values becomes equivalent to the rank which the individual would have obtained through non-dominated comparison.

In Fig. 3, $Si$ represents the array of individuals sorted on each objective $i$, $position(p, Si)$ returns the position of $p$ in $Si$, $sort(P)$ returns a standard sort of $P$ based on objective value, $p_j(obj)$ represents the objective value of individual $p_j$ and $set\_position(p_j, Qi, pos)$ will set the position of $p_j$ as $pos$ in $Qi$.

We estimate the runtime complexity of the above procedure as follows: the $sort\_on\_objective$ procedure can be completed in $O(N \log_2 N)$. This has to be repeated for M objectives, hence the overall complexity is $O(MN \log_2 N)$.

The arrays sorted on objectives have a size of $O(N)$. But we will need M such arrays and hence the space complexity increases to $O(MN)$, which can be quite a desirable trade-off even for moderate sizes of M and N.

## 5 THE CLASSIFICATION OF TYPES OF LEUKEMIA

Here we present the multiobjective problem to which we apply the algorithms described above. The task for the optimizer is to identify the optimal gene subsets which classify microarray gene expression data. Both the algorithms were run in similar conditions on the 50-gene Leukemia dataset of Golub et al. [8]. This dataset contains the expression data of 7129 human genes taken from 38 samples of patients, 27 of whom were suffering from Acute lymphoblastic leukemia (ALL) and 11 of whom were suffering from Acute myeloid leukemia (AML).

We adopt three different conflicting objectives: to minimize the number of genes used in classification while maintaining acceptable classification accuracy expressed

$fast\_non\_dominated\_sort(P)$
{
  for each $p \in P$
    $S_p = \phi$
    $n_p = 0$
    for each $q \in Q$
      if $(p \prec q)$ then
        $S_p = S_p \cup \{q\}$
      else if $(q \prec p)$ then
        $n_p = n_p + 1$
    if $n_p = 0$ then
      $p_{rank} = 1$
      $F_1 = F_1 \cup \{p\}$
$i = 1$
  while $F_i \neq \phi$
    $Q = \phi$
    for each $p \in F_i$
      for each $q \in S_p$
        $n_q = n_q - 1$
        if $n_q = 0$ then
          $q_{rank} = i + 1$
          $Q = Q \cup \{q\}$
$i = i + 1$
$F_i = Q$
}

Fig. 2. Non-dominated sorting procedure adopted in NSGA-II [4].

as training error and testing error [14].

The multi-objective optimization problem is formulated as follows:
1) The first objective function - The gene subset identification task is to minimize the number genes in a subset or to minimize the gene subset size for a classifier.
2) The second objective function - Minimize the number of class prediction mismatches in the training samples. These are calculated using the Leave-Out-One-Cross-Validation (LOOCV).
3) The third objective function - To minimize the number of class prediction mismatches in the test samples, these are calculated using the classifier constructed based on all samples in the training set.

We have used a weighted-voting approach to predict the class of a sample based on such informative gene subsets and a set of samples with known class labels, and the LOOCV procedure is used to determine the number of mismatches in the training samples. Thereafter, the classifier is constructed using such informative gene subsets and all the training samples. Then, the performance of the classifier is estimated using the remaining samples in a test-set.

The NSGA-II described above is used for handling the above three conflicting objectives. During each generation, the algorithm finds gene subsets which classify the data into the two sub-types of leukemia. As the evolutionary algorithm proceeds, progressively better gene subsets are identified.

## 6 RESULTS AND DISCUSSION

We carried out a comparative study between the basic algorithm (NSGA-II) [4], the variant due to Jensen (NSGA-IIb) [7] and our own variant NSGA-IIa. The optimization was carried for gene subset selection as described in section 5 above. Tests were carried out on a Pentium-IV, 2.0 GHz machine with 1 GB RAM, running Fedora Core 6.0 operating system.

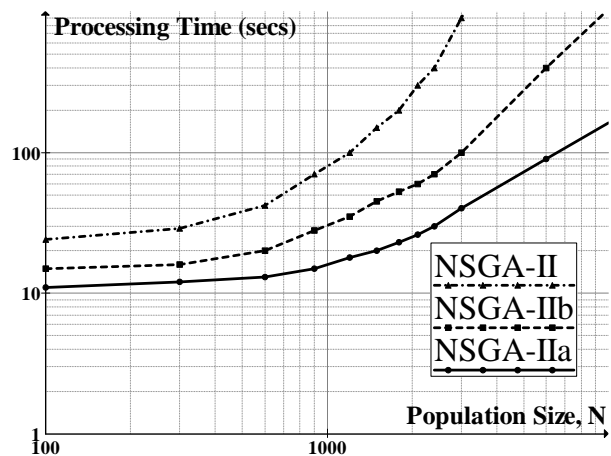

Fig. 4. Graph showing the variation of processing time in seconds with the population size, N.

Fig. 4 shows the variation of average processing time for convergence, in CPU seconds, taken over ten runs in each case when the population size is gradually increased, in steps. It is evident from the graph that the NSGA-IIa we propose is faster than the NSGA-IIb proposed by Jensen. The basic NSGA-II performs even worse since its complexity is $O(MN^2)$. Though the difference is negligible for small populations, there is a marked difference for population sizes of 1000 and more. This is significant since such problems frequently demand the use of large population sizes in order to yield reasonable results.

## 7 CONCLUSION

In this work we have proposed a variant of the NSGA-II algorithm which has a lower runtime complexity. The lower complexity is achieved by trading off space against time in the non-dominated sorting stage.

A comparative study between the basic NSGA II, and a variant of this due to Jensen and our improved algorithm shows that our algorithm performs better when population sizes are larger than 1000. We are currently working on applying this improved algorithm to other standard test problems, as well as other datasets.

$faster\_non\_dominated\_sort(P)$

{

  $for\ each\ objective\ i$

    $S_i = sort\_on\_objective(P, i)$

  $for\ each\ p\ in\ P$

    $p_{rank} = \sum_{i=1}^{M} position(p, S_i)$

}

$sort\_on\_objective(P, i)$

{

  $pos = 0$

  $Q_i = sort(P)$

  $for\ each\ p_j\ in\ P$

    $if\ (p_j(obj) \neq p_{j-1}(obj))$

      $pos = pos + 1$

    $set\_position(p_j, Q_i, pos)$

  $return\ Q_i$

}

Fig. 3. A faster non-dominated sorting procedure as compared to the one of NSGA-II.

## ACKNOWLEDGMENT

This paper is an extended version of our paper "A Time-Efficient Variant of the Non-Dominated Sorting Genetic Algorithm," which has been accepted for oral presentation at the First IFIP International Conference on Bioinformatics, to be held at Surat, India, from 25th to 28th March, 2010.

## REFERENCES

[1] J. Andersson, "A Survey of Multiobjective Optimization in Engineering Design," Technical report LiTH-IKP-R-1097, Dept of Mechanical Engg, Linkping University, Sweden, 2000, pp. 34.

[2] J. Knowles and D. Corne, "The Pareto archived evolution strategy: A new baseline algorithm for multiobjective optimization," in Proceedings of the 1999 Congress on Evolutionary Computation. Piscataway, NJ: IEEE Press, 1999, pp. 98–105.

[3] E. Zitzler and L. Thiele, "Multiobjective optimization using evolutionary algorithms—A comparative case study," in Parallel Problem Solving From Nature, V, A. E. Eiben, T. Bäck, M. Schoenauer, and H.-P. Schwefel, Eds. Berlin, Germany: Springer-Verlag, 1998, pp. 292–301.

[4] K. Deb, A. Pratap, S. Agarwal, and T. Meyarivan, "A Fast and Elitist Multi-objective Genetic Algorithm: NSGA-II," IEEE Trans. Evol. Comp., vol. 6, no. 2, Apr. 2002, pp. 182-197.

[5] H. Fang, Q. Wang, Y. Tu, & M.F. Horstemeye, "An efficient non-dominated sorting method for evolutionary algorithms," IEEE Trans. on Evolutionary Computation, Vol. 16, Issue 3, Fall 2008, pp. 355-384, 2008.

[6] K. D. Tran, "An Improved Non-dominated Sorting Genetic Algorithm-II (ANSGA-II) with adaptable parameters," Intl. Jour. of Intelligent Systems Technologies and Applications, Vol. 7, No. 4, Sept 2009, pp. 347-369(23), Inderscience, 2009.

[7] M. T. Jensen, "Reducing the run-time complexity of multi-objective EAs: The NSGA-II and other algorithms," IEEE Transactions on Evolutionary Computation, 7:502-515, 2003.

[8] T.R. Golub, D.K. Slonim, P. Tamayo, C. Huard, M. Gaasenbeek, J.P. Mesirov, H. Coller, M.L. Loh, J.R. Downing, M.A. Caligiuri, C.D. Bloomfield, and E.S. Lander, "Molecular Classification of Cancer: Class Discovery and Class Prediction by Gene Expression Monitoring," Science, vol. 286, 1999, pp. 531-537.

[9] D. Parrott, L. Xiaodong, V. Ciesielski, "Multi-objective techniques in genetic programming for evolving classifiers," The 2005 IEEE Congress on Evolutionary Computation, Vol. 2, pp. 1141-1148, 2005.

[10] K. Deb, Multi-objective Optimization using Evolutionary Algorithms, Wiley, Chichester, UK, 2001.

[11] C. M. Fonseca and P. J. Fleming, "Genetic algorithms for multiobjective optimization: Formulation, discussion and generalization," in Proceedings of the Fifth International Conference on Genetic Algorithms, S. Forrest, Ed. San Mateo, CA: Morgan Kauffman, 1993, pp. 416–423.

[12] N. Srinivas and K. Deb, "Multiobjective function optimization using nondominated sorting genetic algorithms," Evol. Comput., vol. 2, no. 3, pp. 221–248, Fall 1995.

[13] P. K. Kumar, Sharath S., R. G. D'Souza, K. Chandra Sekaran, "Memetic NSGA—A Multi-Objective Genetic Algorithm for Classification of Microarray Data," adcom, pp. 75-80, 15th International Conference on Advanced Computing and Communications (ADCOM 2007), IEEE Computer Society, 2007.

[14] K. Deb and A.R. Reddy, "Classification of Two-class Cancer Data Reliably Using Evolutionary Algorithms," Publ. of Kanpur Genetic Algorithms Lab., India, KanGAL Report No. 2003001, 2003.

**Rio G. L. D'Souza** is a Research Scholar at the Department of Computer Engineering at National Institute of Technology Karnataka, India. He is currently on sabbatical leave from St Joseph Engineering College, Mangalore. His research interests include Soft Computing, Computer Networks, and Bioinformatics. He is a member of IEEE and IEEE Computational Intelligence Society.

**Dr. K. Chandra Sekaran** is a Professor of Computer Engineering at National Institute of Technology Karnataka, India. His research includes Computer Networks, Dependable Network / Distributed computing, Autonomic computing and Community Informatics. He has 20 years of teaching and research and one year Industry experience. He has published more than 86 publications in International and National proceedings and authored two books. He was the Organizing Chair of 14th International Conference ADCOM 2006, International Symposium on Ad Hoc and Ubiquitous Computing ISAHUC'06. He has served as a member of Program Committee in various International conferences and also reviewer in many Journals. He has supervised sponsored projects and IT consultant to some corporates in this region of India.

**Dr. A. Kandasamy** is a Professor at the Department of Mathematical and Computational Sciences, National Institute of Technology Karnataka, India. In the recent past, he has been a Visiting Faculty at IES, School of Engineering & Technology, Asian Institute of Technology, Thailand. His research experience spans over 21 years and he has a teaching experience of more than 16 years. His research interests include Computational Techniques and Algorithms, Computational Fluid Dynamics, Optimization Techniques, Stochastic Processes, Genetic Algorithms.